\documentclass[12pt]{article}

\usepackage{amsmath,amsfonts,epsfig,mathrsfs,yfonts}
\usepackage{color}
\usepackage{xcolor}
\usepackage{cite}

\usepackage{stmaryrd}
\SetSymbolFont{stmry}{bold}{U}{stmry}{m}{n}

\usepackage{mdframed}
\usepackage{scalerel}
\usepackage[top=2.5cm, bottom=2.5cm, left=2.5cm, right=2.5cm]{geometry} 
\usepackage{dsfont}
\numberwithin{equation}{section}
\usepackage{hyperref}

\hypersetup{
	colorlinks=true,
	linkcolor=black,
	filecolor=blue,
	citecolor=blue,      
	urlcolor=cyan,
}
\usepackage[most]{tcolorbox}
\usepackage{float}
\usepackage{tablefootnote}

\usepackage{url}

\usepackage{mathtools}
\usepackage{float,array,booktabs,multirow}
\usepackage{enumitem}
\usepackage{graphicx,shortvrb}
\usepackage{newlfont}
\usepackage{tensor}

\newcommand{\ra}[1]{\renewcommand{\arraystretch}{#1}}

\usepackage{caption}
\def\wd{\wedge}
\def\w{\omega}

\def\d{\text{d}}

\usepackage{fancyhdr} 
\newcommand{\auth}{\large Onur Ayberk \c{C}akmak ${}^{a}$\footnote{email: acakmak@metu.edu.tr, \;\; ORCID ID: 0009-0005-0555-8294}
	and {\"O}zg{\"u}r Sar{\i}o\u{g}lu ${}^a$\footnote{email: sarioglu@metu.edu.tr, \;\; ORCID ID: 0000-0003-2282-3510}}

\begin{document}
	
	
	\begin{center}
		{\Large{\bf Chern-Simons potentials of higher-dimensional Pontryagin densities}}
		\vspace{1cm}
		
		\auth
	\end{center}
	\vspace{.5cm}
	\centerline{${}^a${\it \small Department of Physics, Faculty of Arts and Sciences,}}
	\centerline{{\it \small Middle East Technical University, 06800, Ankara, T{\"u}rkiye}}
	
	\date{\today}
	
	\vspace{1cm}
	
	\centerline{{\bf Abstract}}
	We develop a novel and systematic approach to computing the 
	$(2n-1)$-form Chern-Simons potential given the Pontryagin density, i.e. the $n^{\text{th}}$ Chern character, in arbitrary even dimensions \( D=2 n \geq 2 \). Throughout we work with a generic affine connection, that results in a non-vanishing torsion in general, and allows for non-metricity, which accommodates the existence of non-trivial Chern characters and hence Pontryagin densities. We outline an algorithm, with its implementation 
	as a code, which lets one to determine the Chern-Simons potential 
	given the Pontryagin density in an arbitrary even dimension.
	
	\bigskip
	
	
	\vspace{10cm}
	
	
	\bigskip
	
	\small
	\pagebreak
	\tableofcontents

	\renewcommand{\thefootnote}{\arabic{footnote}}
	\setcounter{footnote}{0}
	
	\section{Introduction} \label{sec:intro}
	Topological invariants and their corresponding currents often 
	play a critical role in the understanding of quantum field theories, 
	modified gravitational models, and condensed matter systems in 
	contemporary theoretical physics. In four spacetime dimensions,
	two fundamental objects in this context are the Pontryagin density 
	(PD) and the associated Chern-Simons (CS) potential. 
	These emerge naturally in field theories with gauge and diffeomorphism invariance, and 
	have important consequences in various areas such as cancellation of the infamous ABJ anomaly, which arises from the calculation of 
	tadpole chiral current \cite{Bell:1969ts,Adler:1969gk}, 
	topological phases of matter \cite{Zee:1995avy,Fujita:2009kw}, 
	three-dimensional topologically massive gravity 
	\cite{Deser:1981wh,Deser:1982vy,Alexander:2009tp},
	modifications of general relativity, particularly in CS modified gravity \cite{Jackiw:2003pm}, 
	and play a central role in the construction of instanton solutions, e.g. of the Yang-Mills theory 
	\cite{Belavin:1975fg}. Moreover, the relationship between PDs and CS potentials is central to anomaly inflow mechanisms and plays a role 
	in the bulk-boundary correspondence \cite{Callan:1984sa,Davoyan:2023zhi}. From the perspective of mathematics, the closely 
	related Pontryagin classes are used in the classification of fiber bundles \cite{Nakahara:2003nw}.
	
	PD is a topological term constructed from the curvature of a given gauge or affine connection. 
	It has mostly been utilized in the discussion of four-dimensional gauge theories. It follows from the definition of the \emph{total
		Pontryagin class} \cite{Nakahara:2003nw}
	\begin{equation}\label{totPc}
		\mathcal{P}(F) := \det{\Big( I + \frac{F}{2 \pi} \Big)} \,,
	\end{equation}
	where \( F := \d A + A \wd A \) is the 2-form field strength of the 1-form gauge field $A$.
	\( \mathcal{P}(F) \) or its geometric counterpart 
	\( \mathcal{P}(R) \) can readily be expanded as a finite sum over the \emph{Pontryagin classes}, e.g. in $D=4$
	\begin{equation}\label{sumPcs}
		\mathcal{P}(R) = 1 + p_{1}(R) + p_{2}(R) \,,
	\end{equation}
	where \( p_{k}(R) \, (k=1, 2) \) is a polynomial of order
	$2k$, and is explicitly given as \cite{Nakahara:2003nw}
	\begin{equation}\label{eq:Pcs12}
		\begin{aligned}
			p_{1}(R) =& - \frac{1}{8 \pi^2} \text{Tr}(R\wd R) =
			- \frac{1}{8 \pi^2} R^{ab} \wd R_{ba} \,, \\
			p_{2}(R) =& \frac{1}{128 \pi^4} \Big( 
			\big(\text{Tr}(R\wd R) \big)^2 -2 \, \text{Tr}(R\wd R \wd R \wd R) \Big) =
			\frac{1}{16 \pi^4} \det R \,.
		\end{aligned}
	\end{equation}
	Obviously, \( p_{2}(R) \) identically vanishes as a 
	differential form in \( D=4 \).
	
	On the other hand, there are also the \emph{Chern characters} which
	are defined as \cite{Nakahara:2003nw}
	\begin{equation}\label{defChs}
		\text{ch}_{k}(F) := \frac{1}{k!} \text{Tr}\Big( \frac{\text{i} F}{2 \pi} \Big)^k \,, \quad \big( k=0, 1, \dots, [D/2] \big)
	\end{equation}
	and in $D=4$ one has
	\begin{equation}\label{2ndChs}
		\text{ch}_{2}(R) = - \frac{1}{8 \pi^2} \text{Tr}(R\wd R) \,,
	\end{equation}
	which exactly coincides with \( p_{1}(R) \) defined in \eqref{eq:Pcs12}. This specific coincidence that occurs only in 
	$D=4$ has led to a somewhat regrettable misnomer in the physics
	literature ever since the publication of the seminal works \cite{Deser:1981wh,Deser:1982vy,Jackiw:2003pm}: The phrase 
	\emph{Pontryagin density}\footnote{See e.g. the paragraph containing eqn. (3.17) of \cite{Deser:1981wh}.} (PD)\footnote{Since \( \text{ch}_{n}(R)\) by itself is a \emph{tensor density}, and \emph{not} a tensor that requires any \emph{weighting}, it can be integrated directly 
		in the construction of proper physical actions as in e.g. \cite{Nakagawa:2020gqc,Lindstrom:2022qtr}; hence the ``D" in ``PD". To the best of our knowledge, we are not aware of any use of the concatenation of the terms ``Pontryagin" and ``density" in the mathematics literature.} is specifically (but somewhat erroneously) used for denoting the $n^{\text{th}}$ Chern character \( \text{ch}_{n}(R)\)
	in an arbitrary even dimension \( D=2 n \geq 2 \). In what follows, we conspicuously continue using the phrase ``Pontryagin density" instead of
	the absolutely precise but cumbrous ``$n^{\text{th}}$ Chern character".
	
	It is well known that the integral of Pontryagin class over a compact, 
	oriented 4-manifold gives the Pontryagin number and describes a topological invariant \cite{PontNo}. It is a direct consequence of this property that the so-called instanton number is proportional to the integral of the PD both in the field theoretical 
	and gravitational contexts and yields also the soliton solutions \cite{Hawking:1976jb,Rajaraman:1982is}. Moreover, it can be written as a total derivative and does 
	not affect the local dynamics of a theory when integrated over a compact space without a boundary. 
	However, it leads to non-trivial effects in the presence of boundaries or when other fields are 
	coupled to itself.
	
	The CS potential is intimately related to the PD and arises naturally when seeking to express 
	it as a total divergence \cite{Nakahara:2003nw}. The four-dimensional PD satisfies 
	\begin{equation}\label{eq:P4}
		\d \, \text{ch}_{2}(F) = 0 \,.
	\end{equation}
	By Poincar\'{e}'s lemma, \eqref{eq:P4} suggests that this PD can be locally 
	written as the derivative of a 3-form 
	\begin{equation}\label{eq:P4_CS4}
		\text{ch}_{2}(F) = \d \, {\mathcal K}_3(A) \,,
	\end{equation}
	where the CS potential ${\mathcal K}_3$ is given by
	\begin{equation}\label{eq:CS3}
		{\mathcal K}_3(A) = \text{Tr} \left( A \wd \d A + \dfrac{2}{3}A \wd A \wd A \right) \,. 
	\end{equation}
	Obviously, an analogous relation holds for the geometric counterpart $\text{ch}_{2}(R)$,
	that guarantees the conservation of topological charge, where now the gauge field $A$ 
	in \eqref{eq:CS3} is replaced by the corresponding connection 1-form $\omega$.
	
	The integral of \eqref{eq:CS3} on a given three-dimensional 
	manifold is topological, i.e. it doesn't depend on the metric of 
	the manifold but only on its topology, and is the ubiquitous 
	three-dimensional CS action of topologically massive gauge theory 
	as well as its gravitational counterpart 
	\cite{Deser:1981wh,Deser:1982vy}. This theory is a prime 
	example of a topological quantum field theory 
	\cite{Witten:1988ze,Witten:1988hc}. Finally, we must  
	mention that due to the rich structure provided by CS theory in 
	the description of three-dimensional invariants, it also plays 
	an important role in knot theory, via Witten's work 
	\cite{Witten:1988hf} relating it to the Jones polynomial 
	\cite{Jones:1985dw}.
	
	The special role that the dimension $D$ plays in the definition 
	of the PD needs to be recalled as well. In the original setting, 
	it is obvious that Pontryagin classes containing an \emph{odd 
		number} of wedged ``$F$"s or ``$R$"s vanish due to the underlying symmetry of \eqref{totPc} \cite{Nakahara:2003nw}. 
	However, it is possible to have non-vanishing PDs, i.e. 
	\( \text{ch}_{n}(R) \) in \( D = 2n \geq 6 \) using a 
	\emph{generic affine connection}. Considering the Einstein-Hilbert term together with the 
	relevant PD coupled to, e.g., a scalar field $\theta$ 
	\[ S = \int d^{2n}x \, \left( \sqrt{-g} \, R + \theta \, \text{ch}_{n}(R) \right) \,, \]
	as in the case of Chern-Simons modified gravity in $D=4$ \cite{Jackiw:2003pm}, or further
	performing a (consistent) circle (or sphere) truncation (compactification) of the action $S$, may have interesting 
	physical and geometrical consequences that is worth looking into. One, of course, first needs 
	to determine the explicit form of the CS potential given the PD in an even dimension 
	$D=2n \geq 6$.
	
	Thus the present work grew out of a desire to explore the 
	mathematical structure of the PD and its associated CS potential 
	in gauge and gravitational theories in arbitrary even dimensions.
	While trying to better understand this relationship, we have 
	recognized that this is explicitly investigated for relatively 
	low dimensions from \( D=4 \) to \( D=8 \) only 
	\cite{Lindstrom:2022qtr,Radu:2020yea}. This work basically 
	develops a novel and systematic approach to computing the CS 
	potentials given the PDs in arbitrary higher even dimensions, and
	explicitly extends the basic ideas laid out in Subsection 11.5.2 of Nakahara \cite{Nakahara:2003nw}.
	
	\section{Preliminaries}
	Let us start by introducing the notation we use in this work. Throughout the section, we work on a $D=2n$ dimensional spacetime manifold $M$ with $n \geq 2$. From now on, we suppress the local Lorentz indices in our calculations and discard the explicit mentioning of tracing `Tr' in all expressions. However, they can be easily reinserted by contracting the second index of each connection with the first index of the next one since the cyclic order of the $\d\w$ and 
	$\w$ pieces remains unchanged\footnote{An explicit example of how this works can be found in appendix \ref{sec:appb}.}
	\[
	\d\tensor{\w}{^{i_1}_{i_2}} \wd \d\tensor{\w}{^{i_2}_{i_3}}\wd \dots \wd \tensor{\w}{^{i_n}_{i_1}}. 
	\]
	Moreover, we use 
	\begin{equation}\label{eq:wk}
		(\d\w)^k := \; \underbrace{(\d\w)\wd(\d\w)\wd\dots\wd(\d\w)}_{\text{$k$ times}} \,, \qquad
		\w^l := \; \underbrace{\w\wd\w\dots\wd\w}_{\text{$l$ times}} \,.
	\end{equation}
	The PD in $2n$-dimensions is given by\footnote{The coefficient in front is chosen to normalize the coefficient of the $(\d\w)^n$ piece.} 
	\begin{equation}\label{eq:P2n_exp}
		\begin{aligned}
			\Big( \frac{2 \pi}{\text{i}} \Big)^{n} \, n! \, \text{ch}_{n}(R) =: \mathcal{P}_{2n} =&\; \text{Tr} (R^n) :=\; \text{Tr} \big( \underbrace{R\wd\dots \wd R}_{\text{$n$ times}} \big) \\
			=&\; 
			\left(
			\d\w + \w^2
			\right)\wd  \dots \wd 
			\left(
			\d\w + \w^2 
			\right) \\
			=&\; (\d\w)^n + \d\w\wd\w^{2n-2} + \w^2\wd \d\w\wd \w^{2n-4} + \dots + \w^{2n} \\
			=&\; (\d\w)^n + n\;\d\w\wd\w^{2n-2} + n\; (\d\w)^2\wd\w^{2n-4} + \dots + \w^{2n},
		\end{aligned}
	\end{equation}
	where the final term `$\w^{2n}$' vanishes due to cyclic symmetries, but we include it anyway for the sake of completeness. Furthermore, we use the properties of the wedge product to sum the terms that are equal to each other. 
	From the last line of \eqref{eq:P2n_exp}, we note that each piece contains terms in $\w$ wedged an even number of times. 
	Thus, each term appearing on the right-hand-side (RHS) of \eqref{eq:P2n_exp} is a \emph{permutation} of $a$ number of `$\d\w$'s and $(n-a)$ number of `$\w^{2}$'s with \(a=0, 1, \dots, n\).
	However, once we apply integration-by-parts (IBP) to `$\d\w$'s, terms with $\w$ wedged an odd number of times will be generated. To distinguish these, we shall refer to the terms as \emph{odd} or \emph{even} depending on whether they contain any $\w^{\text{odd}}$ or not
	\begin{equation}
		\bigwedge\limits_k(\d\w)^{A_k}\wd\w^{B_k} \text{ is an }
		\begin{cases}
			\text{\emph{even} term,} & \text{when } B_k \text{ is even for all } k, \\
			\text{\emph{odd} term,} & \text{otherwise}.
		\end{cases}
	\end{equation} 
	We can classify the even terms by the total number of `$\d\w$'s and `$\w^2$'s they contain. To this end, we define the \emph{permutation sets} (that partition terms appearing in \eqref{eq:P2n_exp})\footnote{We use this as an opportunity to fix an error in equation (A.22) of \cite{tez}.}
	\begin{equation}\label{eq:Pset}
		P^{\{\bar{a},n-a\}} := 
		\left\{
		\bigwedge_k (\d\w)^{A_k}\wd\w^{2B_k} \Big| \sum\limits_{k}A_k=a, \;\sum\limits_k B_k = n-a
		\right\},
	\end{equation} 
	where $A_k, B_k$ and $a$ are non-negative integers with 
	$0 \leq a \leq n$. The anti-symmetric nature of the PD enables us to establish equivalence relations. We call two permutations equivalent and denote this by $\stackrel{\text{cyc}}{\sim}$, if they can be transformed into each other by applying an even number of cyclic shifts. Mathematically, this corresponds to e.g. expressing the equivalence of the following
	\begin{equation}\label{eq:equivalence}
		\begin{aligned}
			(\d\w)^{A_1}\wd\w^{2B_1}\wd(\d\w)^{A_2}\wd\w^{2B_2}\wd \dots \stackrel{\text{cyc}}{\sim}&\; (\d\w)^{A_1-1}\wd\w^{2B_1}\wd(\d\w)^{A_2}\wd\w^{2B_2}\wd\dots \wd \d\w \\
			\stackrel{\text{cyc}}{\sim}&\; (\d\w)^{A_1-2}\wd \w^{2B_1}\wd (\d\w)^{A_2}\wd\w^{2B_2}\wd \dots \wd (\d\w)^2 \\
			& \hspace{3cm} \vdots \\
			\stackrel{\text{cyc}}{\sim}&\; \w^{2B_1}\wd(\d\w)^{A_2}\wd\w^{2B_2}\wd \dots \wd (\d\w)^{A_1} \\
			\stackrel{\text{cyc}}{\sim}&\; \w^{2B_1-2}\wd(\d\w)^{A_2}\wd\w^{2B_2}\wd\dots \wd(\d\w)^{A_1}\wd\w^2 \\
			& \hspace{3cm} \vdots 
		\end{aligned}
	\end{equation} 
	We refer the reader to \eqref{eq:app_equivalence} for a similar calculation with restored Lorentz indices. Clearly, $P^{\{\bar{a},n-a\}}$ can be partitioned into equivalence classes by the relation $\stackrel{\text{cyc}}{\sim}$. The cyclic equivalence class of a permutation $\bigwedge_k (\d\w)^{A_k}\wd\w^{2B_k}\in P^{\{\bar{a},n-a\}}$ can then be constructed as 
	\begin{equation}
		\begin{aligned}
			\left[
			\bigwedge\limits_k (\d\w)^{A_k}\wd\w^{2B_k}
			\right]_{\text{cyc}} = 
			\left\{
			x\in P^{\{\bar{a},n-a\}} \Big| \; x \stackrel{\text{cyc}}{\sim} \bigwedge\limits_k (\d\w)^{A_k}\wd\w^{2B_k}
			\right\}.
		\end{aligned}
	\end{equation}
	Thus, instead of working with the whole of $P^{\{\bar{a},n-a\}}$, we pick one random element `$x_a$' from each equivalence class and construct a new set of representatives 
	\begin{equation}\label{eq:P_dist}
		S^{\{\bar{a},n-a\}}_\text{dist} := 
		\left\{
		\text{One element $x_a$ from each $P^{\{\bar{a},n-a\}}/\stackrel{\text{cyc}}{\sim}$, $a=1,2,\dots,n$} 
		\right\},
	\end{equation}
	which we refer to as \emph{distinct permutation sets}. As an illustrative example, $P^{\{\bar{2},2\}}$ is given by
	\begin{equation}\label{eq:P22}
		\begin{aligned}
			P^{\{\bar{2},2\}} &= 
			\big\{
			(\d\w)^2\wd\w^4, (\d\w)\wd\w^4\wd(\d\w), \w^4\wd(\d\w)^2, \w^2\wd(\d\w)^2\wd\w^2, \\
			& \qquad (\d\w)\wd\w^2\wd(\d\w)\wd\w^2, \w^2\wd(\d\w)\wd\w^2\wd(\d\w)
			\big\},
		\end{aligned}
	\end{equation}
	which contains six elements in total. However, the first four 
	can be transformed into each other by means of even number of cyclic shifts through the wedge product, which, separately, is also the case for the last two elements. Thus, four of the elements of \eqref{eq:P22} are in fact redundant. We remove these equivalent terms by constructing 
	$S^{\{\bar{2},2\}}_{\text{dist}}$ using the prescription \eqref{eq:P_dist}
	\begin{equation}
		S^{\{\bar{2},2\}}_{\text{dist}} = 
		\left\{
		(\d\w)^2\wd\w^4, (\d\w)\wd\w^2\wd(\d\w)\wd\w^2
		\right\},
	\end{equation}
	and the equivalence classes of its elements are 
	\begin{equation}
		\begin{aligned}
			&\left[
			(\d\w)^2\wd \w^4 
			\right]_{\text{cyc}} =\; 
			\big\{
			(\d\w)^2\wd\w^4, (\d\w)\wd\w^4\wd(\d\w), \w^4\wd(\d\w)^2, \w^2\wd(\d\w)^2\wd\w^2
			\big\}, \\
			&\left[
			(\d\w)\wd\w^2\wd(\d\w)\wd\w^2
			\right]_{\text{cyc}} =\; 
			\big\{
			(\d\w)\wd\w^2\wd(\d\w)\wd\w^2, \w^2\wd(\d\w)\wd\w^2\wd(\d\w)
			\big\}.
		\end{aligned}
	\end{equation}
	Using these, the PD can be rewritten as 
	\begin{equation}\label{eq:P2n}
		\begin{aligned}
			\mathcal{P}_{2n} =&\; \sum\limits_{a=1}^{n}\;\sum\limits_{i=1}^{\left|P^{\{\bar{a},n-a\}}\right|} p_i, \qquad p_i\in P^{\{\bar{a},n-a\}} \\
			=&\; \sum\limits_{a=1}^{n}\;\sum\limits_{i=1}^{\left|S_{\text{dist}}^{\{\bar{a},n-a\}}\right|} \Big| [s_i]_{\text{cyc}}\Big| s_i, \qquad s_i\in S^{\{\bar{a},n-a\}}_{\text{dist}},
		\end{aligned}
	\end{equation}
	where strokes $\left| \, \right|$ denote the number of elements in a set. Finally, the symbol `$\stackrel{i,j}{=}$' is used to indicate the following: The exterior derivative `$\d$' is applied to the $j$th $\w$ piece within the $i$th derivative group. For instance, in
	\begin{equation}
		\begin{aligned} 
			(\d\w)\wd\w^2\wd(\d\w)^3\wd\w^4 \stackrel{2,2}{=}&\; \d 
			\left(
			(\d\w)\wd\w^2\wd(\d\w)\wd
			\overset{\downarrow}{\w} \wd(\d\w)\wd\w^4
			\right) \\
			&- (\d\w)\wd \d(\w^2)\wd(\d\w)\wd\w\wd(\d\w)\wd\w^4 \\
			&+(\d\w)\wd\w^2\wd(\d\w)\wd\w\wd(\d\w)\wd\d(\w^4),
		\end{aligned} 
	\end{equation}
	the first `2' refers to the group $(\d\w)^3$ whereas the second `2' specifies the second `$\d\w$' within $(\d\w)^3$.
	
	\section{Calculation of CS potentials}
	In this section, we introduce our method for deriving CS potentials from higher-dimensional PDs. Our approach follows the notation presented in \cite{tez}, with the distinction that we employ differential forms for clarity and conciseness. 
	
	For ease of discussion, we start by reproducing the expansion \eqref{eq:P2n_exp} here
	\begin{equation}\label{eq:P2n_exp2}
		\begin{aligned}
			\mathcal{P}_{2n} =&\; R^n:=\underbrace{R\wd\dots \wd R}_{\text{$n$ times}}  
			= \; 
			\left(
			\d\w + \w^2
			\right)\wd  \dots \wd 
			\left(
			\d\w + \w^2 
			\right) \\
			=&\; (\d\w)^n + n\;\d\w\wd\w^{2n-2} + n\; (\d\w)^2\wd\w^{2n-4} + \dots + \w^{2n}.
		\end{aligned}
	\end{equation}
	The coefficients of the permutation terms of the expansion in \eqref{eq:P2n_exp2} are given by the minimum even number of cyclic shifts which leave the term invariant. As an example, the even term 
	\[
	(\d\w)\wd\w^4\wd(\d\w)\wd\w^4,
	\]
	that contributes to $\mathcal{P}_{12}$ is left invariant under three double-cyclic rotations (meaning a total of six rotations) and hence its coefficient in the expansion is 3. On the other hand, for the term
	\[
	(\d\w)^2\wd\w^8
	\]
	a full cyclic permutation (that is six double-cyclic shifts) is needed, which leads to a coefficient of 6 in the expansion. The coefficients of the even terms resulting in the expansion of 
	the PD for \( 2 \leq n \leq 8\) are given in Table \ref{tab:coeffTable} without referring to the terms themselves.
	
	\begin{table}[H]\centering
		\ra{1.1}
		\begin{tabular}{@{}cccccccccc@{}}
			\toprule
			$D=2n$ & $(\d\w)^0$ & $(\d\w)^1$ & $(\d\w)^2$ & $(\d\w)^3$ & $(\d\w)^4$  & $(\d\w)^5$ & $(\d\w)^6$ & $(\d\w)^7$ & $(\d\w)^8$  \\
			\midrule
			4  & 1 & 2 & 1 & & & & & &\\
			\hline                            
			6 & 1 & 3 & 3 & 1 & & & & &\\
			\hline                            
			8  & 1 & 4 & 4\:\:2 & 4 & 1 & & & &\\
			\hline                            
			10 & 1 & 5 & $5^2$ & $5^2$ & 5 & 1 & & &\\
			\hline                            
			12 & 1 & 6 & $6^2$\:\:3 & $6^3$\:\:2 & $6^2$\:\:3 & 6 & 1 & &\\
			\hline                            
			14 & 1 & 7 & $7^3$ & $7^5$ & $7^5$ & $7^3$ & 7 & 1 & \\
			\hline                            
			16 & 1 & 8 & $8^3$\:\:4 & $8^7$ & $8^8$\:\:4\:\:2 & $8^7$ & $8^3\:\:4$ & 8 & 1 \\
			\bottomrule
		\end{tabular}
		\caption{The coefficients of the even terms in the expansion of the PD for \( 2 \leq n \leq 8\).}
		\label{tab:coeffTable}
	\end{table}
	Here, each column represents a different permutation set \eqref{eq:Pset} and the exponents in the first row correspond to the total number of `$\d\w$' pieces of the corresponding sets, which is denoted by `$\bar{a}$' in the previous section. Each number in a cell is the coefficient of an even permutation. For convenience, repeated coefficients are indicated by exponents. For instance, in $D=12$ the contribution from the elements of 
	$P^{\{\bar{3},3\}}$ 
	to the expansion \eqref{eq:P2n_exp2} is
	\[
	6(\dots) + 6(\dots) + 6(\dots) + 2(\dots),
	\]
	where the terms in the parentheses are all elements of 
	$S^{\{\bar{3},3\}}_\text{dist}$. The number of elements of a permutation set $P^{\{\bar{a},n-a\}}$ equals the sum of all the coefficients in the corresponding cell in Table \ref{tab:coeffTable}. Consequently, in each row for $n$, the coefficients add up to $2^{n}$. For instance, in $D=12$ using the convention mentioned, we get 
	\[
	(1) + (6) + (2\times6+3) + (3\times6+2) + (2\times6+3) + (6) + (1) = 64 = 2^{6},
	\]
	where we have split the contribution from different sets \eqref{eq:Pset} using parentheses. Note that the sums of the coefficients in 
	the parentheses are
	\[
	1 \quad 6 \quad 15 \quad 20 \quad 15 \quad 6 \quad 1
	\]
	and these follow from the binomial expansion of $(\d\w+\w^2)^6$, which naturally holds for all $D$ as well. The coefficients for 
	the $4\leq D \leq 32$ cases have been calculated via a Matlab 
	\cite{matlab2023b} code, the output of which can be found 
	in \cite{code}. 
	
	Applying IBP to a `$\d\w$' term in a permutation from \eqref{eq:P2n_exp2}, say $s\in S^{\{\bar{a},n-a\}}_\text{dist}$, yields an equation like 
	\begin{equation}
		s = \d(\dots) + \text{even terms} + \text{odd terms},
	\end{equation}
	where all even terms belong to the same set 
	$P^{\{\bar{a},n-a\}}$ (and $S^{\{\bar{a},n-a\}}_{\text{dist}}$ after cyclic shifting), since the number of total `$\d\w$'s and `$\w^2$'s are the same in these terms.  Performing IBP to each `$\d\w$' in each element of $S^{\{\bar{a},n-a\}}_\text{dist}$ yields a maximum number of 
	\( \widetilde{m}^{\{\bar{a},n-a\}}
	:=a\times \left| S^{\{\bar{a},n-a\}}_\text{dist} \right| \) 
	equations. However, it is possible for two different applications of IBP to generate the exact same equation, even though they may be arranged differently. In fact, equations that include the same total derivative term typically result in the same final equation. For instance,
	\begin{equation}
		\begin{aligned}
			&\underbrace{(\d\w)^3\wd \w^4}_{e_1} \stackrel{1,3}{=} \; \d 
			\left(
			(\d\w)^2\wd \w^5
			\right)+ \underbrace{(\d\w)^2\wd\w\wd(\d\w)\wd\w^3}_{o_1} - \underbrace{(\d\w)^2\wd\w^2\wd(\d\w)\wd\w^2}_{e_2} \\
			&\hspace{2.3cm} + \underbrace{(\d\w)^2\wd\w^3\wd(\d\w)\wd\w}_{o_2} - \underbrace{(\d\w)^2\wd\w^4\wd(\d\w)}_{=e_1} \,, \\
			&\underbrace{(\d\w)^2\wd\w^2\wd(\d\w)\wd\w^2}_{=e_2} \stackrel{2,1}{=}\; \d
			\left( 
			(\d\w)^2\wd\w^5
			\right) - \underbrace{(\d\w)^3\wd\w^4}_{=e_1} + \underbrace{(\d\w)^2\wd\w\wd(\d\w)\wd\w^3}_{=o_1} \\
			&\hspace{2.3cm} + \underbrace{(\d\w)^2\wd\w^3\wd(\d\w)\wd\w}_{=o_2} - \underbrace{(\d\w)^2\wd\w^4\wd(\d\w)}_{=e_1}
		\end{aligned}
	\end{equation}
	are two copies of the same equality. Consequently, a large number of the equations are, in fact, redundant. We are interested in the remaining independent 
	\( m^{\{\bar{a},n-a\}} \leq \widetilde{m}^{\{\bar{a},n-a\}} \) equations which produce distinct total derivative terms. In Table \ref{tab:mTable}, we highlight how the number of redundant equations increases with the dimension of the space.
	
	\begin{table}[H]\centering
		\ra{1.1}
		\begin{tabular}{@{}cccccc@{}}
			\toprule
			$D=2n$ & \( (\bar{a},n-a) \) & \( \left| S^{\{\bar{a},n-a\}}_\text{dist} \right| \) & 
			\( \widetilde{m}^{\{\bar{a},n-a\}} \) & \( m^{\{\bar{a},n-a\}} \) & \# of red. eqns\\
			\midrule
			\multirow{1}{*}{4}  & $(\bar{1},1)$ & 1 & 1 & 1 & 0 \\
			\hline
			\multirow{2}{*}{6}  & $(\bar{1},2)$ & 1 & 1 & 1 & 0 \\
			& $(\bar{2},1)$ & 1 & 2 & 1 & 1 \\
			\hline                            
			\multirow{3}{*}{8}  & $(\bar{1},3)$ & 1 & 1 & 1 & 0 \\
			& $(\bar{2},2)$ & 2 & 4 & 1 & 3 \\
			& $(\bar{3},1)$ & 1 & 3 & 2 & 1 \\
			\hline
			\multirow{4}{*}{10} & $(\bar{1},4)$ & 1 & 1 & 1 & 0 \\
			& $(\bar{2},3)$ & 2 & 4 & 1 & 3 \\
			& $(\bar{3},2)$ & 2 & 6 & 3 & 3 \\
			& $(\bar{4},1)$ & 1 & 4 & 2 & 2 \\
			\hline
			\multirow{5}{*}{12} & $(\bar{1},5)$ & 1 & 1 & 1 & 0\\
			& $(\bar{2},4)$ & 3 & 6 & 1 & 5\\
			& $(\bar{3},3)$ & 4 & 12 & 4 & 8\\
			& $(\bar{4},2)$ & 3 & 12 & \hspace{7pt}4\,\tablefootnote{
				The number of independent total derivative terms for this configuration seems to be 5. However, there is an additional relation between four of them given in \eqref{eq:deriv_rel}. Hence, the real number of independent terms is reduced to 4.} & 8 \\
			& $(\bar{5},1)$ & 1 & 5 & 3 & 2 \\
			\bottomrule
		\end{tabular}
		\caption{The calculation of \( \widetilde{m}^{\{\bar{a},n-a\}} \),  \( m^{\{\bar{a},n-a\}} \) and the number 
			of redundant equations for distinct permutation sets from $D=4$ to $D=12$.}
		\label{tab:mTable}
	\end{table}
	
	Next, we cast the $m^{\{\bar{a},n-a\}}$ equations into a matrix equation of the form
	\begin{equation}\label{eq:matrix_eqn}
		\text{M}^{\{\bar{a},n-a\}}_\text{even} \text{u}^{\{\bar{a},n-a\}}_\text{even} = \d \text{u}^{\{\bar{a},n-a\}}_\d + \text{M}^{\{\bar{a},n-a\}}_\text{odd} \text{u}^{\{\bar{a},n-a\}}_\text{odd},
	\end{equation}
	where 
	\begin{itemize}[left=0pt]
		\item{$\text{M}^{\{\bar{a},n-a\}}_\text{even}$ is an $m^{\{\bar{a},n-a\}}\times \left| S^{\{\bar{a},n-a\}}_\text{dist} \right|$-dimensional matrix containing the coefficients of the even terms,}
		\item{$\text{M}^{\{\bar{a},n-a\}}_\text{odd}$ is an $m^{\{\bar{a},n-a\}}\times$(\# of odd terms)-dimensional matrix containing the coefficients of the odd terms,}
		\item{$\text{u}^{\{\bar{a},n-a\}}_\text{even}$ is a 
			$\left| S^{\{\bar{a},n-a\}}_\text{dist} \right|$-dimensional column vector containing the even terms,}
		\item{$\text{u}^{\{\bar{a},n-a\}}_\d$ is an $m^{\{\bar{a},n-a\}}$-dimensional column vector containing the total derivative terms,}
		\item{$\text{u}^{\{\bar{a},n-a\}}_\text{odd}$ is a (\# of odd terms)-dimensional column vector containing the odd terms.}
	\end{itemize}
	For a generic dimension and permutation set, it is hard to predict $m^{\{\bar{a},n-a\}}$ and the emergent odd terms. However, $\text{u}^{\{\bar{a},n-a\}}_\text{even}$ will be of the form
	\begin{equation}
		\text{u}^{\{\bar{a},n-a\}}_\text{even} = 
		\begin{pmatrix}
			k_1 (\d\w)^a \wd \w^{2(n-a)} \\
			k_2(\d\w)^{a-1}\wd \w^2\wd (\d\w) \wd \w^{2(n-a-1)} \\
			\vdots 
		\end{pmatrix},
	\end{equation}
	where all elements of $S^{\{\bar{a},n-a\}}_\text{dist}$ are contained in the column vector and $k_i$'s are the dimensions of the cyclic equivalence class ($[\cdot]_\text{cyc}$) of the related element in the row. For instance, 
	\begin{equation*}
		k_1 := \left\lvert [(\d\w)^a \wd \w^{2(n-a)}]_\text{cyc} \right\rvert, \quad
		k_2 := \left\lvert [(\d\w)^{a-1}\wd \w^2\wd (\d\w) \wd \w^{2(n-a-1)}]_\text{cyc} \right\rvert,
	\end{equation*}
	and so on.
	
	The aim of the whole process is to eliminate the odd terms, which do not appear in \eqref{eq:P2n_exp2}, from the equality \eqref{eq:matrix_eqn}. To accomplish this, we look for a constant coefficient matrix $\text{K}^{\{\bar{a},n-a\}}$ satisfying
	\begin{equation}\label{eq:Modd0}
		\text{K}^{\{\bar{a},n-a\}} \, 
		\text{M}^{\{\bar{a},n-a\}}_\text{odd} = 0.
	\end{equation}
	We want to project into a proper nontrivial subspace that will allow us to further kill the second term in the RHS of \eqref{eq:matrix_eqn}. Note that $\text{K}^{\{\bar{a},n-a\}}$ becomes a row-vector when all redundancies and dependent equations are completely eliminated. Multiplying \eqref{eq:matrix_eqn} by a proper $\text{K}^{\{\bar{a},n-a\}}$ satisfying \eqref{eq:Modd0}, one will be left with
	\begin{equation}\label{eq:Keven_part}
		\text{K}^{\{\bar{a},n-a\}} \, 
		\text{M}^{\{\bar{a},n-a\}}_\text{even} \, 
		\text{u}^{\{\bar{a},n-a\}}_\text{even} = \d 
		\left(
		\text{K}^{\{\bar{a},n-a\}} \, 
		\text{u}_\d^{\{\bar{a},n-a\}}
		\right) \,.
	\end{equation}
	From the onset, Poincar\'e's lemma guarantees that the PD can be written as a total derivative. Hence, \emph{the row-vector} $\text{K}^{\{\bar{a},n-a\}}$ will be of the form 
	\begin{equation}\label{eq:K}
		\text{K}^{\{\bar{a},n-a\}} \,
		\text{M}^{\{\bar{a},n-a\}}_\text{even} = 
		k^{\{\bar{a},n-a\}} \, (1\;1\;\dots\;1) \,,
	\end{equation}
	where $k^{\{\bar{a},n-a\}}$ is a constant to be determined for each $a$, so that the left hand side of \eqref{eq:Keven_part} generates the contribution of $S^{\{\bar{a},n-a\}}_\text{dist}$ to the PD
	\begin{equation}\label{eq:Ku_even}
		\text{K}^{\{\bar{a},n-a\}} \, 
		\text{M}^{\{\bar{a},n-a\}}_\text{even} \, 
		\text{u}^{\{\bar{a},n-a\}}_\text{even} = k^{\{\bar{a},n-a\}} \, \sum\limits_{i=1}^{{
				\left|
				S^{\{\bar{a},n-a\}}_\text{dist}
				\right|}} 
		\left| 
		[s_i]_\text{cyc}
		\right| \, s_i \,, \qquad 
		s_i\in S^{\{\bar{a},n-a\}}_\text{dist} \,.
	\end{equation}
	When there are redundant terms, which may be caused by the dependence of two total derivative terms, then $\text{K}^{\{\bar{a},n-a\}}$ and, hence, $\text{K}^{\{\bar{a},n-a\}} \, \text{M}^{\{\bar{a},n-a\}}_\text{even}$ contain multiple rows. Again, as ensured by Poincaré's lemma, at least one of the rows takes the form in \eqref{eq:K}. The other rows yield subtle relations between the dependent total derivative terms. As an example in $D=12$, \eqref{eq:Keven_part} with $a=4$ is given by\footnote{See section B.4.2 of \cite{tez}.}
	\begin{equation}\label{eq:D12_ex}
		\begin{aligned}
			\begin{pmatrix}
				8 & 8 & 8 \\
				0 & 0 & 0
			\end{pmatrix}&
			\begin{pmatrix}
				6 (\d\w)^4\wd\w^4 \\
				6(\d\w)^3\wd\w^2\wd(\d\w)\wd\w^2 \\
				3(\d\w)^2\wd\w^2\wd(\d\w)^2\wd\w^2
			\end{pmatrix} =& \\
			&\;\; \d 
			\left[
			\begin{pmatrix}
				3 & 3 & -1 & 4 & 1 & 0 \\
				0 & -1 & 1 & -1 & 0 & 1
			\end{pmatrix}
			\begin{pmatrix}
				6\w\wd(\d\w)^3\wd\w^4 \\
				6\w\wd(\d\w)^2\wd\w^4\wd(\d\w) \\
				6\w\wd(\d\w)\wd\w^4\wd(\d\w)^2 \\
				6\w\wd(\d\w)^2\wd\w^2\wd(\d\w)\wd\w^2 \\
				6\w\wd(\d\w)\wd\w^2\wd(\d\w)\wd\w^2\wd(\d\w) \\
				6\w\wd(\d\w)\wd\w^2\wd(\d\w)^2\wd\w^2
			\end{pmatrix}
			\right].
		\end{aligned}
	\end{equation}
	The first row of \eqref{eq:D12_ex} is exactly eight times $ \text{the contribution from } S^{\{\bar{4},2\}}_\text{dist} = \d (\dots)$, while the second row picks out the relation 
	\begin{equation}\label{eq:deriv_rel}
		\begin{aligned}
			&- \d\left[\w\wd(\d\w)^2\wd\w^4\wd(\d\w)\right] + \d\left[\w\wd(\d\w)\wd\w^4\wd(\d\w)^2\right] \\ 
			&\qquad - \d\left[\w\wd(\d\w)^2\wd\w^2\wd(\d\w)\wd\w^2\right] + \d\left[\w\wd(\d\w)\wd\w^2\wd(\d\w)^2\wd\w^2\right] = 0.
		\end{aligned}
	\end{equation}
	This equality can be verified by applying IBP a second time to either of the elements and making use of the fact that $\d^2 = 0$. Combining the contributions from different $S^{\{\bar{a},n-a\}}_\text{dist}$ sets, $(a=1,2,\dots, n)$, and making use of \eqref{eq:P2n}, \eqref{eq:Keven_part}, \eqref{eq:Ku_even} yields
	\begin{equation}\label{eq:P2n_d}
		\begin{aligned}
			\mathcal{P}_{2n} &= \d
			\left(
			\sum\limits_{a=1}^n \, \dfrac{1}{k^{\{\bar{a},n-a\}}} \, \text{K}^{\{\bar{a},n-a\}} \, 
			\text{u}^{\{\bar{a},n-a\}}_\d 
			\right).
		\end{aligned}
	\end{equation}
	The sum inside the brackets in \eqref{eq:P2n_d} gives the higher dimensional CS potential
	\begin{equation}\label{eq:CS}
		\mathcal{K}_{2n-1} := \sum\limits_{a=1}^n \,
		\dfrac{1}{k^{\{\bar{a},n-a\}}} \, \text{K}^{\{\bar{a},n-a\}} \, \text{u}^{\{\bar{a},n-a\}}_\d \,.
	\end{equation}
	The CS potentials up to and including dimension $D=12$ have been calculated explicitly in \cite{tez}. We summarize the results in Table \ref{tab:summary}. 
	
	\begin{table}[H]\centering
		\ra{2}
		\begin{tabular}{@{}cl@{}}
			\toprule
			$D$ & CS Potential \\
			\hline 
			\multirow{1}{*}{4} & $\w\wd
			\left( 
			(\d\w)+\dfrac{2}{3}\w^2
			\right)$ \\
			\hline
			\multirow{1}{*}{6} & $\w\wd
			\left( 
			(\d\w)^2 + \dfrac{3}{2}(\d\w)\wd\w^2 + \dfrac{3}{5}\w^4
			\right)$ \\
			\hline 
			\multirow{1}{*}{8} & $\w\wd
			\left( 
			(\d\w)^3 + \dfrac{8}{5}(\d\w)^2\wd\w^2 + \dfrac{4}{5}(\d\w)\wd\w^2\wd(\d\w) + 2(\d\w)\wd\w^4 + \dfrac{4}{7}\w^6
			\right)$ \\
			\hline 
			\multirow{2}{*}{10} & $\w\wd
			\bigg( 
			(\d\w)^4 + \dfrac{5}{3}(\d\w)^3\wd\w^2 + \dfrac{5}{3}(\d\w)^2\wd\w^2\wd(\d\w) +  \dfrac{15}{7}(\d\w)^2\wd\w^4$ \\ 
			& $\hspace{1cm} + \dfrac{5}{7}(\d\w)\wd\w^4\wd(\d\w) + \dfrac{10}{7}(\d\w)\wd\w^2\wd(\d\w)\wd\w^2 +\dfrac{5}{2}(\d\w)\wd\w^6 + \dfrac{5}{9}\w^8
			\bigg)$ \\
			\hline 
			\multirow{4}{*}{12} & $\w\wd
			\bigg(
			(\d\w)^5 + \dfrac{12}{7}(\d\w)^4\wd\w^2 + \dfrac{12}{7}(\d\w)^3\wd\w^2\wd(\d\w) + \dfrac{6}{7}(\d\w)^2\wd\w^2\wd(\d\w)^2$ \\
			& $\hspace{1cm} + \dfrac{9}{4}(\d\w)^3\wd\w^4 + \dfrac{9}{4}(\d\w)^2\wd\w^4\wd(\d\w) - \dfrac{3}{4}(\d\w)\wd\w^4\wd(\d\w)^2$ \\
			& $\hspace{1cm} + 3(\d\w)^2\wd\w^2\wd(\d\w)\wd\w^2 + \dfrac{3}{4}(\d\w)\wd\w^2\wd(\d\w)\wd\w^2\wd(\d\w) + \dfrac{8}{3}(\d\w)^2\wd\w^6$ \\
			& $\hspace{1cm}  + \dfrac{2}{3}(\d\w)\wd\w^6\wd(\d\w) + 2(\d\w)\wd\w^5\wd(\d\w)\wd\w + \dfrac{4}{3}(\d\w)\wd\w^3\wd(\d\w)\wd\w^3$ \\
			& $\hspace{1cm} + 3(\d\w)\wd\w^8 + \dfrac{6}{11}\w^{10} 
			\bigg)$ \\
			\bottomrule
		\end{tabular}
		\caption{CS potentials from $D=4$ to $D=12$.}
		\label{tab:summary}
	\end{table}
	
	However, for $D\geq 14$ the CS potentials are much more complicated since both the number of equations and the number of odd terms increase drastically. As an indication for the complexity of the computation, here we present the total number of equations (after eliminating the redundant ones) to be solved from $D=14$ to $D=32$ in Table \ref{tab:eqnum}.
	\begin{table}[H]
		\centering
		\begin{tabular}{c|cccccccccc}
			$D$ & 14 & 16 & 18 & 20 & 22 & 24 & 26 & 28 & 30 & 32 \\
			\hline
			\# of equations & 24 & 49 & 87 & 165 & 302 & 598 & 1081 & 2137 & 3954 & 7513
		\end{tabular}
		\caption{The total number of equations to be solved from $D=14$ to $D=32$.}
		\label{tab:eqnum}
	\end{table}
	\noindent 
	Hence, one has to resort to computational methods, the algorithm of which we outline in appendix \ref{sec:app}. As an illustration of how the algorithm works, we also present the explicit calculation in $D=8$ in appendix \ref{sec:appb}.
	
	\section{Discussion}
	In this work, we have developed a systematic way of calculating the CS potential of a given PD 
	in arbitrary even dimensions $D \geq 4$. The procedure we have put forward consists of 
	\begin{itemize}
		\item{expanding the corresponding PD of the given even dimension,} 
		\item{grouping the permutations based on the number of $\d\w$ terms they contain,}
		\item{applying IBP on each of those $\d\w$ pieces,}
		\item{summing the equivalent terms by making use of the cyclic permutation property,}
		\item{excluding the redundant equations and solving the matrix equation generated by 
			different IBP to find the contribution to the total derivative term (The explicit number of 
			redundant equations from $D=4$ to $D=12$ for different distinct permutation sets is given 
			in Table \ref{tab:mTable}),}
		\item{adding up all the contributions and writing the PD as a total derivative to arrive at the corresponding CS potential.}
	\end{itemize}   
	
	In lower even dimensions $D \leq 8$, the calculation can be performed with relative ease. 
	We hereby provide the CS potentials computed from the PDs in $D=4$ to $D=12$ in Table 
	\ref{tab:summary}. However, in higher even dimensions $D \geq 14$, the computation 
	becomes much more complicated due to the increasing number of $\d\w$ pieces. To this 
	end, we present an algorithm in appendix \ref{sec:app} for the calculation of CS potentials 
	from PDs. Our algorithm is implemented as a Matlab \cite{matlab2023b} code and is accessible in GitHub \cite{code}. We have run our code for \( 4 \leq D \leq 32 \), the explicit 
	output\footnote{We have explicitly checked the correctness of the code's output with our 
		hand-made calculations for \( 4 \leq D \leq 12 \).} of which can also be found there.
	
	As already mentioned in section \ref{sec:intro}, the calculation
	of CS potentials from higher even-dimensional PDs has limited 
	coverage in the literature, and we were able to find (beyond the 
	$D=4$ case) only the $D=6$ and $D=8$ cases being addressed 
	\cite{Lindstrom:2022qtr,Nakahara:2003nw,Radu:2020yea}. This work fixes this shortcoming in the literature. As implicitly 
	assumed in our derivations, we have throughout worked with a 
	generic affine connection that results in general a non-trivial
	torsion and non-metricity that allows for the existence of 
	non-vanishing PDs in \( D = 4n-2 \; (n \geq 2) \) dimensions. 
	It also goes without saying that the CS potentials thus determined
	are defined up to the addition of an exact $(D-1)$-form.
	
	We hope to return to the physical implications of our 
	computations with an emphasis on their role in topologically 
	non-trivial configurations and modified theories of gravity in 
	higher dimensions.
	
	\bigskip
	
	\noindent{\bf Acknowledgments}
	
	\noindent
	We would like to thank Rasim Y{\i}lmaz for a careful reading of this manuscript, and anonymous referees for helping us forge this paper into a more resilient form.
	
	\bigskip
	
	\noindent{\bf Data Availability}
	
	\noindent
	As mentioned in the manuscript, the implementation of the algorithm 
	developed in this work as a Matlab code is accessible in GitHub 
	(https://github.com/CaesarSilvae/pontryagin-calc.git) and is licensed 
	under CC BY 4.0.
	
	\bigskip
	
	\noindent{\bf Conflict of interest}
	
	\noindent
	The corresponding author states that there is no conflict of interest.
	
	\bigskip
	\appendix
	\section{The computation in \texorpdfstring{$D=8$}{D=8}}
	\label{sec:appb}
	In this appendix we give an illustrative example on how the algorithm works, and explicitly calculate the PD 
	\( \mathcal{P}_8 \) in $D=8$ .
	
	In $D=8$, the expansion \eqref{eq:P2n_exp} (suppressing the trace operation ``Tr") reads
	\begin{equation}\label{eq:app_P8}
		\begin{aligned}
			\mathcal{P}_8 & =  (\d\w)^4 + (\d\w)^3\wd\w^2 + (\d\w)^2\wd\w^2\wd(\d\w) + (\d\w)\wd\w^2\wd(\d\w)^2 + \w^2\wd(\d\w)^3 \\
			& \quad + (\d\w)^2\wd\w^4 + (\d\w)\wd\w^2\wd(\d\w)\wd\w^2 + (\d\w)\wd\w^4\wd(\d\w) + \w^2\wd(\d\w)^2\wd\w^2 \\
			& \quad + \w^2\wd(\d\w)\wd\w^2\wd(\d\w) + \w^4\wd(\d\w)^2 + (\d\w)\wd\w^6 + \w^2\wd(\d\w)\wd\w^4 \\
			& \quad + \w^4\wd(\d\w)\wd\w^2 + \w^6\wd(\d\w) \,, \\
			& = (\d\w)^4 + 4(\d\w)^3\wd\w^2 + 4(\d\w)^2\wd\w^4 + 2(\d\w)\wd\w^2\wd(\d\w)\wd\w^2 + 4(\d\w)\wd\w^6,
		\end{aligned}
	\end{equation}
	where we have discarded the vanishing $\w^8$ term. The sets defined
	in \eqref{eq:Pset} read 
	\begin{equation*}
		\begin{aligned}
			P^{\{\bar{4},0\}} &= 
			\big\{
			(\d\w)^4
			\big\}, \\
			P^{\{\bar{3},1\}} &= 
			\big\{
			(\d\w)^3\wd\w^2,(\d\w)^2\wd\w^2\wd(\d\w),(\d\w)\wd\w^2\wd(\d\w)^2,\w^2\wd(\d\w)^3
			\big\}, \\
			P^{\{\bar{2},2\}} &= 
			\big\{
			(\d\w)^2\wd\w^4,(\d\w)\wd\w^2\wd(\d\w)\wd\w^2,(\d\w)\wd\w^4\wd(\d\w), \\ 
			&\qquad \w^2\wd(\d\w)^2\wd\w^2,w^2\wd(\d\w)\wd\w^2\wd(\d\w),\w^4\wd(\d\w)^2
			\big\}, \\
			P^{\{\bar{1},3\}} &= 
			\big\{
			(\d\w)\wd\w^6,\w^2\wd(\d\w)\wd\w^4,w^4\wd(\d\w)\wd\w^2,\w^6\wd(\d\w)
			\big\} \,,
		\end{aligned}
	\end{equation*}
	and the distinct partition sets \eqref{eq:P_dist} are 
	\begin{equation*}
		\begin{aligned}
			S_\text{dist}^{\{\bar{4},0\}} &= \left\{(\d\w)^4\right\}, \\
			S_\text{dist}^{\{\bar{3},1\}} &= \left\{(\d\w)^3\wd\w^2\right\}, \\
			S_\text{dist}^{\{\bar{2},2\}} &= \left\{(\d\w)^2\wd\w^4,(\d\w)\wd\w^2\wd(\d\w)\wd\w^2\right\}, \\
			S_\text{dist}^{\{\bar{1},3\}} &= \left\{(\d\w)\wd\w^6\right\}.
		\end{aligned}
	\end{equation*}
	As an illustration of the redundancy mentioned in the text, we show that the elements of $P^{\{\bar{3},1\}}$ are indeed equivalent under cyclic shifts. Restoring the Lorentz indices of the first member, we have 
	\begin{equation}\label{eq:app_equivalence}
		\begin{aligned}
			\d\tensor{\w}{^{i_1}_{i_2}}\wd \d\tensor{\w}{^{i_2}_{i_3}}\wd \d\tensor{\w}{^{i_3}_{i_4}}\wd \tensor{\w}{^{i_4}_{i_5}}\wd \tensor{\w}{^{i_5}_{i_1}} &= \d\tensor{\w}{^{i_2}_{i_3}} \wd \d\tensor{\w}{^{i_3}_{i_4}}\wd \tensor{\w}{^{i_4}_{i_5}}\wd \tensor{\w}{^{i_5}_{i_1}} \wd \d\tensor{\w}{^{i_1}_{i_2}}, \\
			&= \d\tensor{\w}{^{i_3}_{i_4}}\wd \tensor{\w}{^{i_4}_{i_5}}\wd \tensor{\w}{^{i_5}_{i_1}} \wd \d\tensor{\w}{^{i_1}_{i_2}} \wd \d\tensor{\w}{^{i_2}_{i_3}}, \\
			&= \tensor{\w}{^{i_4}_{i_5}}\wd \tensor{\w}{^{i_5}_{i_1}} \wd \d\tensor{\w}{^{i_1}_{i_2}} \wd \d\tensor{\w}{^{i_2}_{i_3}} \wd \d\tensor{\w}{^{i_3}_{i_4}},
		\end{aligned}
	\end{equation}
	where we used the property of the wedge product to shift the leftmost $\d\w$ piece to the end in each line. The RHS of the first line gives the second member of $P^{\{\bar{3},1\}}$ once we rename the indices $i_5\to i_4\to i_3\to i_2 \to i_1\to i_5$. Similarly, the second line yields the third member with $i_5\to i_3\to i_1\to i_4\to i_2\to i_5$, and the third line yields the fourth member with $i_5\to i_2\to i_4\to i_1 \to i_3\to i_5$.
	
	In what follows, we apply IBP to the distinct elements, where the
	$\w$ on which IBP is applied is written in bold.
	
	\subsection{\texorpdfstring{$\{\bar{4},0\}$}{\{ \={4},0 \}}}
	Since there is no $\w$ piece, IBP applied on any $\d\w$ piece directly generates the total derivative term
	\begin{equation}\label{eq:app_4-0}
		\begin{aligned}
			(\d\w)^4 &\stackrel{1,1}{=} \d[\boldsymbol{\w}\wd(\d\w)^3].
		\end{aligned}
	\end{equation}
	
	\subsection{\texorpdfstring{$\{\bar{3},1\}$}{\{ \={3},1 \}}}
	The two independent equations are
	\begin{equation*}
		\begin{aligned}
			4(\d\w)^3\wd\w^2 &\stackrel{1,1}{=} \d[4\boldsymbol{\w}\wd(\d\w)^2\wd\w^2] + 4\boldsymbol{\w}\wd(\d\w)^3\wd\w - 4\boldsymbol{\w}\wd(\d\w)^2\wd\w\wd(\d\w) \\
			&= \d[4\boldsymbol{\w}\wd(\d\w)^2\wd\w^2] - 4(\d\w)^3\wd\w\wd\boldsymbol{\w} + 4(\d\w)^2\wd\w\wd(\d\w)\wd\boldsymbol{\w} \\
			4(\d\w)^3\wd\w^2 &\stackrel{1,2}{=} \d[4(\d\w)\wd\boldsymbol{\w}\wd(\d\w)\wd\w^2] + 4(\d\w)\wd\boldsymbol{\w}\wd(\d\w)^2\wd\w - 4(\d\w)\wd\boldsymbol{\w} \wd(\d\w)\wd\w\wd(\d\w) \\
			&= \d[4\boldsymbol{\w}\wd(\d\w)\wd\w^2\wd(\d\w)] - 4(\d\w)^2\wd\w\wd(\d\w)\wd\boldsymbol{\w} - 4(\d\w)^2\wd\boldsymbol{\w}\wd(\d\w)\wd\w.
		\end{aligned}
	\end{equation*}
	The IBP on the third $\d\w$ piece is omitted since it yields the exact same equality as the first equation. Dropping the bold notation, the two equalities can then be cast into the form \eqref{eq:matrix_eqn} as
	\begin{equation*}
		\begin{pmatrix}
			2 \\
			1
		\end{pmatrix} 
		\begin{pmatrix}
			4(\d\w)^3\wd\w^2
		\end{pmatrix} = \d
		\begin{pmatrix}
			4\w\wd(\d\w)^2\wd\w^2 \\
			4\w\wd(\d\w)\wd\w^2\wd(\d\w)
		\end{pmatrix} + 
		\begin{pmatrix}
			1 \\
			-2
		\end{pmatrix}
		\begin{pmatrix}
			4(\d\w)^2\wd\w\wd(\d\w)\wd\w
		\end{pmatrix}.
	\end{equation*}
	Clearly, the $\text{K}^{\{\bar{3},1\}}$ matrix described in \eqref{eq:Modd0} should be of the form 
	\[
	\text{K}^{\{\bar{3},1\}} = c
	\begin{pmatrix}
		2 & 1
	\end{pmatrix},
	\]
	where $c$ is an arbitrary real constant. Multiplying the matrix equation by $\text{K}^{\{\bar{3},1\}}$ from the left and rearranging the coefficients yields 
	\begin{equation}\label{eq:app_3-1}
		4(\d\w)^3\wd\w^2 = \d
		\begin{pmatrix}
			\dfrac{8}{5}(\d\w)^2\wd\w^3 + \dfrac{4}{5}(\d\w)\wd\w\wd(\d\w)\wd\w^2 
		\end{pmatrix} \,.
	\end{equation}
	
	\subsection{\texorpdfstring{$\{\bar{2},2\}$}{\{ \={2},2 \}}}
	This time we have a single independent equation
	\begin{equation*}
		\begin{aligned}
			4(\d\w)^2\wd\w^4 &\stackrel{1,1}{=} \d[4\boldsymbol{\w}\wd(\d\w)\wd\w^4] + 4\boldsymbol{\w}\wd(\d\w)^2\wd\w^3 - 4\boldsymbol{\w}\wd(\d\w)\wd\w\wd(\d\w)\wd\w^2 \\
			&\quad + 4\boldsymbol{\w}\wd(\d\w)\wd\w^2\wd(\d\w)\wd\w - 4\boldsymbol{\w}\wd(\d\w)\wd\w^3\wd(\d\w) \\
			&= \d[4\boldsymbol{\w}\wd(\d\w)\wd\w^4] - 4(\d\w)^2\wd\w^3\wd\boldsymbol{\w} + 4(\d\w)\wd\w\wd(\d\w)\wd\w^2\wd\boldsymbol{\w} \\
			&\quad - 4(\d\w)\wd\w^2\wd(\d\w)\wd\w\wd\boldsymbol{\w} - 4(\d\w)\wd\boldsymbol{\w}\wd(\d\w)\wd\w^3.
		\end{aligned}
	\end{equation*}
	Again, other applications of IBP yield a copy of this equation and thus are redundant. Following the same procedure, the matrix equality reads
	\begin{equation*}
		2
		\begin{pmatrix}
			1 & 1
		\end{pmatrix}
		\begin{pmatrix}
			4(\d\w)^2\wd\w^4 \\
			2(\d\w)\wd\w^2\wd(\d\w)\wd\w^2
		\end{pmatrix} = \d
		\begin{pmatrix}
			4\w\wd(\d\w)\wd\w^4
		\end{pmatrix} \,. 
	\end{equation*}
	So we only need to cancel the 2 factor appearing on the left-hand-side (LHS). Hence, we have 
	\[
	\text{K}^{\{\bar{2},2\}} = \dfrac{1}{2} \,,
	\]
	and 
	\begin{equation}\label{eq:app_2-2}
		4(\d\w)^2\wd\w^4 + 2(\d\w)\wd\w^2\wd(\d\w)\wd\w^2 = \d
		\begin{pmatrix}
			2\w\wd(\d\w)\wd\w^4
		\end{pmatrix} \,.
	\end{equation}
	
	\subsection{\texorpdfstring{$\{\bar{1},3\}$}{\{ \={1},3 \}}}
	This time, there is a single $\d\w$ piece with only one possible IBP choice:
	\begin{equation*}
		\begin{aligned}
			4(\d\w)\wd\w^6 &\stackrel{1,1}{=} \d[4\boldsymbol{\w}\wd\w^6] + 4\boldsymbol{\w}\wd(\d\w)\wd\w^5 - 4\boldsymbol{\w}\wd\w\wd(\d\w)\wd\w^4 + 4\boldsymbol{\w}\wd\w^2\wd(\d\w)\wd\w^3 \\
			&\quad -4\boldsymbol{\w}\wd\w^3\wd(\d\w)\wd\w^2 + 4\boldsymbol{\w}\wd\w^4\wd(\d\w)\wd\w - 4\boldsymbol{\w}\wd\w^5\wd(\d\w) \\
			&= \d[4\boldsymbol{\w}\wd\w^6] - 4(\d\w)\wd\w^5\wd\boldsymbol{\w} - 4(\d\w)\wd\w^4\wd\boldsymbol{\w}\wd\w - 4(\d\w)\wd\w^3\wd\boldsymbol{\w}\wd\w^2 \\
			&\quad - 4(\d\w)\wd\w^2\wd\boldsymbol{\w}\wd\w^3 - 4(\d\w)\wd\w\wd\boldsymbol{\w}\wd\w^4 - 4(\d\w)\wd\boldsymbol{\w}\wd\w^5.
		\end{aligned}
	\end{equation*}
	All the elements on the RHS except the total derivative term are equivalent to that on the LHS. Hence, choosing $\text{K}^{\{\bar{1},3\}}=1/7$ and rearranging, we find 
	\begin{equation}\label{eq:app_1-3}
		4(\d\w)\wd\w^6 = \d
		\begin{pmatrix}
			\dfrac{4}{7}\w^7
		\end{pmatrix} \,.
	\end{equation}
	
	\subsection{Finding \texorpdfstring{$\mathcal{P}_8$}{{\mathcal P}_8}}
	Finally, substituting \eqref{eq:app_4-0}-\eqref{eq:app_1-3} in \eqref{eq:app_P8}, we readily find 
	\begin{equation*}
		\mathcal{P}_8 = \d 
		\left[ \w\wd 
		\left( 
		(\d\w)^3 + \dfrac{8}{5}(\d\w)^2\wd\w^2 + \dfrac{4}{5}(\d\w)\wd\w^2\wd(\d\w) + 2(\d\w)\wd\w^4 + \dfrac{4}{7}\w^6 
		\right)
		\right],
	\end{equation*}
	which is the one given in Table \ref{tab:summary}.
	
	\section{Algorithm} \label{sec:app}
	Here, we give the algorithm that helps us compute the CS potentials in even dimensions. This maybe a bit of a paradigm shift but the starter of our algorithm is to use binary digits to represent permutations: `0' for $\d\w$'s, `1' for $\w^2$'s. With this, a permutation term in $D=2n$ is nothing but a binary sequence consisting of $n$ digits. For example, the term $(\d\w)^2\wd\w^4\wd(\d\w)$ that appears in $D=10$ is represented as the binary sequence $00110$. 
	
	The expansion of the PD in a given dimension contains all possible even terms, which corresponds to the union of all the permutation sets \eqref{eq:Pset} from $a=0$ to $a=n$
	\begin{equation}
		\bigcup_{a=0}^n P^{\{\bar{a},n-a\}} \,.
	\end{equation}
	We denote this union set by $E_n$ when its elements are 
	represented as binary sequences 
	\begin{equation}\label{eq:code_E}
		E_n := \{\underset{0}{\underset{\uparrow}{0\dots 00}},\; \underset{1}{\underset{\uparrow}{0\dots 01}},\; \dots\;,\;
		\underset{2^n-1}{\underset{\uparrow}{1\dots 11}}\} \,.
	\end{equation}
	Although the total number of elements of $E_n$ is $2^n$, the very first `$0\dots 00$' and the very last `$1\dots 11$' ones yield trivial total derivative terms. Nevertheless, we include them for the sake of completeness. The first step to calculate the CS potential is to split $E_n$ into its distinct permutation sets $S^{\{\bar{a},n-a\}}_\text{dist}$. The procedure for doing so
	operates by
	\begin{enumerate}
		\item{picking elements from $E_n$ one by one starting from the very first term labeled as 0 in \eqref{eq:code_E},} 
		\item{determining all distinct cyclic permutations of the picked element by shifting its digits,}
		\item{removing all the distinct permutations found in step 2 from $E_n$ to prevent overcounting,}
		\item{repeating steps 1 to 3 all the way up to the last term labelled as $2^n-1$ in \eqref{eq:code_E}.}
	\end{enumerate}
	As an example, let us go over the $D=4$ case for which the set of even terms is given by 
	\begin{equation}
		E_2=\{00,01,10,11\}.
	\end{equation}
	We first pick 00, which does not have any other equivalent cyclic permutation, and generate the binary equivalent of 
	$P^{\{\bar{2},0\}}$ consisting only of 00. Removing 00 from $E_2$, we are left with $\{01,10,11\}$. The next permutation from the list is 01, whose cyclic equivalence class contains 01 and 10, generating the binary equivalent of $P^{\{\bar{1},1\}}$. We remove these two permutations as well, shrinking the starting set down to $\{11\}$. Similar to the step for 00, 11 has no equivalent cyclic permutation other than itself. The last set $P^{\{\bar{0},2\}}$ is thus generated by $\{11\}$. As a result, this
	subroutine has generated all the cyclic equivalence classes in $D=4$ as
	\[ \{00\}\to P^{\{\bar{2},0\}} \,, \quad 
	\{01,10\}\to P^{\{\bar{1},1\}} \quad 
	\mbox{and} \quad \{11\}\to P^{\{\bar{0},2\}} \,.\]
	
	The next step of the algorithm is to apply IBP to the `$\d\w$' pieces (represented as `$0$'s in the binary sequence). Before applying IBP, the `$1$'s (i.e. `$\w^2$'s) in the permutations 
	must be duplicated. This is so since IBP generates odd terms that contain $\w^{\text{odd}}$ pieces. That is, instead of representing `$\w^2$'s with `$1$'s, we represent each `$\w$' with a single `$1$', e.g. 
	\[ 10100110 \longrightarrow 110110011110 \,. \]
	Now the application of IBP works as follows
	\begin{enumerate}
		\item{The `$0$' that corresponds to the `$\d\w$' on which the IBP is to be applied is converted to a `$1$'; the new binary sequence directly gives the contribution to the CS potential. For instance 
			\begin{equation*}
				11\underset{\underset{1,1}{\uparrow}}{0}11\underset{\underset{2,1}{\uparrow}}{0} 
				\stackrel{1,1}{=} \d(\underbrace{11\overset{\downarrow}{1}110}_{\mathclap{\text{cont. to CS pot.}}}) - \dots
			\end{equation*}
		}
		\item{To apply the derivative to the other `$\w$'s and subtract these corrections, each `$1$' is converted to a 
			`$0$' one at a time and the resulting term is subtracted from the total derivative term $\d(\dots)$. This term is added to the list of even terms $u^{\{\bar{a},n-a\}}_\text{even}$ if the term is even (if it contains only runs of `$1$'s of even length), or to the list of odd terms 
			$u^{\{\bar{a},n-a\}}_\text{odd}$ if the term is odd (if there exists a group of consecutive `$1$'s whose length is odd). To concretely show how it goes, let us reconsider the example given in the previous step
			\[
			110110 \stackrel{1,1}{=} \d(11\overset{\downarrow}{1}110) - \underbrace{011110}_{\text{even term}} - \underbrace{101110}_{\text{odd term}} - \underbrace{111010}_{\text{odd term}} - \underbrace{111100}_{\text{even term}} \,.
			\]
		}
		\item{To find and sum up the cyclically equivalent terms, a convention is needed. To this end, our code picks cyclic permutation with the most number of `$0$'s on the left of the sequence and most number of `$1$'s on the right, which corresponds to the binary number that has the minimum decimal value. After cyclic shifting, the sign in front of each permutation should be updated depending on the number of cyclic shifts needed to put the permutation term to the minimum decimal value according to this convention. One should keep in mind that each `$0$' corresponds to a `$\d\w$' piece and has two suppressed indices, while each `$1$' corresponds to a `$\w$' piece and has a single suppressed index. As a result, when we carry a `$0$' from one end of the binary sequence to the other by means of cyclic shifts, it brings no sign change; but when we do the same for a `$1$', it brings a minus sign. To illustrate these, we rewrite $10100$ and $10111$ as 
\[ 10100 \xrightarrow[\text{2 binary shifts to right}]{\text{4 cyclic shifts to right}} 00101 \,, \quad
	10111 \xrightarrow[\text{1 binary shift to left}]{\text{1 cyclic shift to left}} -01111 \,, \]
			where a cyclic shift cycles each suppressed index once. We emphasize that since `$\d$' and `$\w$' within a `$\d\w$' piece cannot be split, `$0$'s should be shifted from one end to the other through double cyclic shifts.}
		\item{IBP is to be applied to each `$\d\w$' piece in each element of $S^{\{\bar{a},n-a\}}_\text{dist}$. If multiple equations yield the same total derivative term, only one should be considered to eliminate any redundancy. Eq. \eqref{eq:Keven_part} should be solved to find the contribution to the total derivative term from 
			$S^{\{\bar{a},n-a\}}_\text{dist}$.}
		\item{Finally, the whole process is to be repeated for each 
			$S^{\{\bar{a},n-a\}}_\text{dist}$ ($a=0,\dots, n$) and the sum of each contribution eventually gives the CS potential \eqref{eq:CS} for the dimension $D=2n$ considered.}
	\end{enumerate}
	
	The implementation of this algorithm as a Matlab  
	\cite{matlab2023b} code is accessible in GitHub \cite{code}. 
	The explicit output of this code for \( 4 \leq D \leq 32 \) 
	is also available there.

\end{document}